\documentclass[twocolumn,amsmath,amssymb,nofootinbib,floatfix]{revtex4}
\usepackage{epsfig}
\usepackage{graphicx}
\usepackage{amsmath,amssymb,psfrag,bm}

\newcommand{\sN}{\partial}
\newcommand{\sC}{\mathcal{C}}

\newcommand{\sM}{\mathcal{M}}
\newcommand{\sZ}{\mathcal{Z}}
\newcommand{\sH}{\mathcal{H}}
\newcommand{\sG}{\mathcal{G}}

\begin{document}

\title{Cavity approach to the spectral density of non-Hermitian sparse matrices}
\author{Tim Rogers}\affiliation{Department of Mathematics, King's College London, Strand, London WC2R 2LS, United Kingdom}\author{Isaac P\'{e}rez Castillo}\affiliation{Department of Mathematics, King's College London, Strand, London WC2R 2LS, United Kingdom}
\begin{abstract}
The spectral densities of ensembles of non-Hermitian sparse random matrices are analysed using the cavity method. We present a set of equations from which the spectral density of a given ensemble can be efficiently and exactly calculated. Within this approach, the generalised Girko's law is recovered easily. We compare our results with direct diagonalisation for a number of random matrix ensembles, finding excellent agreement.
\end{abstract}
\maketitle
\begin{figure}
\includegraphics[width=240pt]{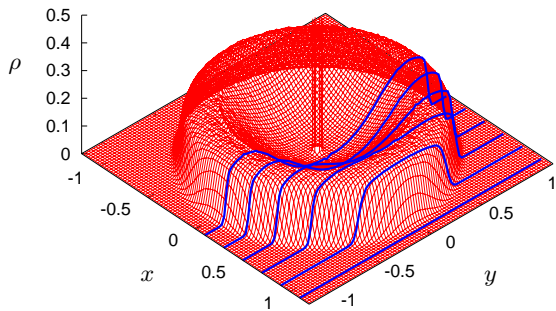}
\begin{picture}(0,0) 
\put(-60,30){$y$}
\put(-190,30){$x$}
\put(-240,110){$\rho$}
\end{picture}
\caption{\label{fig:C2U} Spectral density of asymmetric Poissonian graphs with unitary edge weights and average connectivity $c=2$. The red grid is a histogram of the eigenvalues of $10^5$ samples, the blue lines are the result of the cavity equations, averaged over 1000 samples.}
\end{figure}
\begin{figure}
\includegraphics[width=240pt]{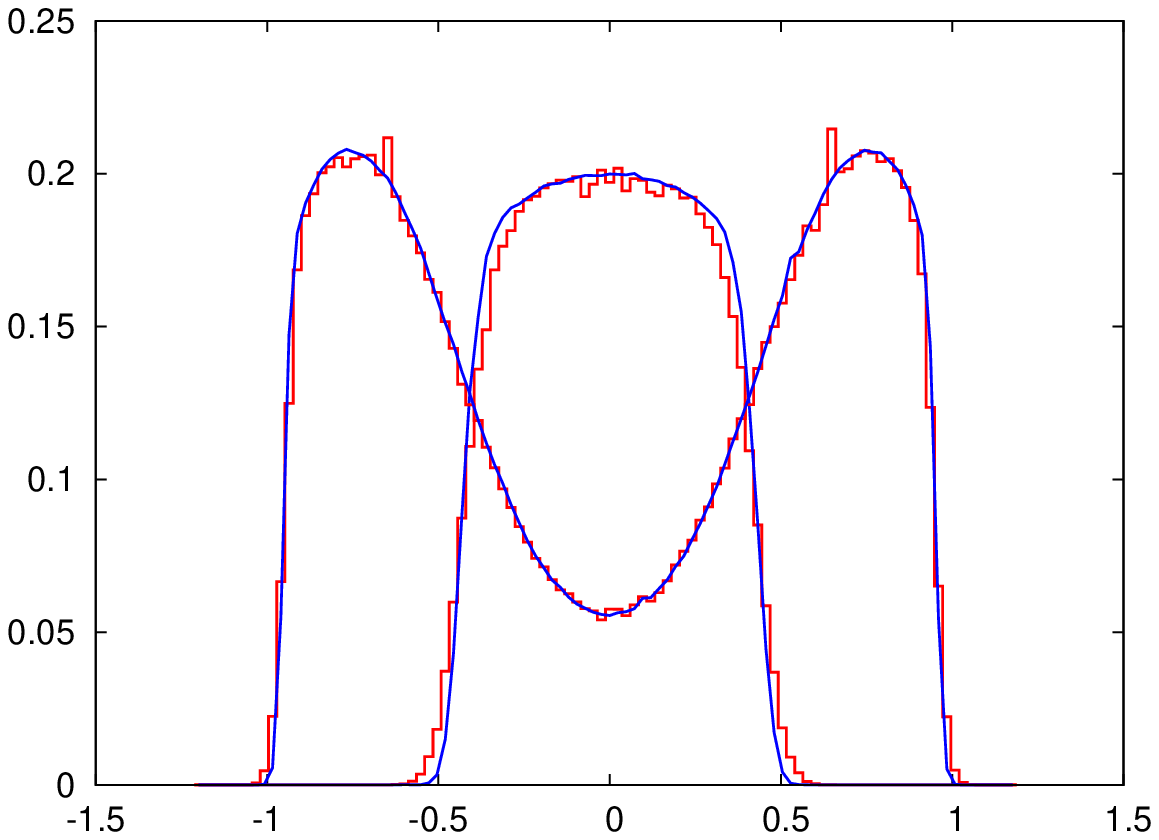}
\begin{picture}(0,0) 
\put(-115,-5){$y$}
\put(-240,100){$\rho$}
\end{picture}
\caption{\label{fig:cuts} Cuts along the lines  $x=0.3$ and $x=0.9$ from Fig. \ref{fig:C2U}}
\end{figure}

For decades, random matrix theory has been the focus of much attention in both physical and mathematical research, with an ever-expanding and remarkably diverse list of applications (for example, see \cite{Guhr1998} for an extensive review of applications in physics). A problem of particular interest is that of determining the spectral density of an ensemble of random matrices. In the early 1950's it was conjuectured that the eigenvalues of certain non-Hermitian random matrix ensembles should be spread evenly throughout the unit disk. Now known as Girko's law, this conjecture has been the subject of many rigorous and non-rigorous studies (e.g. \cite{Girko,Bai1997,tao2007,sommers-1988} and references therein), for various classes of random matrix ensemble.\\
\indent It is a natural desire, then, to extend our understanding of those ensembles that break away from this law. In the more accessible case of real symmetric matrices, it is known that the introduction of sparsity (that is, many entries of the matrix being zero) results in behaviour radically different from that seen in the fully connected limit \cite{rodgersbray,cugliandolo,biroli,nagao}. Sparse real symmetric matrices have been studied extensively, and there exists various approximative schemes \cite{cugliandolo,biroli,nagao}, together with recent exact work \cite{rogers-2008,Reimer,Bordenave}. As we will see, sparsity also has a significant effect on the spectral density of general non-Hermitian matrices, however, this area has not received the attention it deserves, and consequently a great deal remains unknown. \\
\indent In this letter, we tackle the problem of computing the spectral density of sparse non-Hermitian matrices using the cavity method \cite{rogers-2008,MPV,mezard}. A simple closed set of equations is uncovered, whose solution characterises the spectral density of a given matrix. These equations are solved analytically in the fully connected limit, recovering the generalised Girko's law  of \cite{sommers-1988}. We also present the results of a belief-propagation algorithm to quickly compute the solution for ensembles of large sparse matrices.\\
\indent\textit{Preparation.--} Consider an ensemble $\sM$ of $N\times N$ complex, non-Hermitian sparse random matrices. For a given matrix $A\in\sM$, we denote the collection of eigenvalues of $A$ by $\left\{\lambda^A_i: i=1,...,N\right\}$, if $A$ is non-Hermitian, it follows that these $\lambda^A_i$ are complex. For a point $z=x+iy$ in the complex plane, we write the spectral density of $A$ at $z$ as
\begin{equation}
\varrho_A(z,\overline{z})=\frac{1}{N}\sum_{i=1}^N \delta(x-\textrm{Re}\lambda^A_i)\delta(y-\textrm{Im}\lambda^A_i).
\label{rhoA}
\end{equation}
The spectral density of the ensemble, denoted as $\rho(z,\overline{z})$, results from averaging $\varrho_A(z,\overline{z})$ over $\sM$. Following, for example, \cite{sommers-1988,fyodorov-1997, feinberg-1997}, we are able to write\footnote{We use the notation conventions $$\partial_z=\frac{1}{2}\left(\frac{\partial}{\partial x}-i\frac{\partial}{\partial y} \right),\hspace{2mm} \partial_{\overline{z}}=\frac{1}{2}\left(\frac{\partial}{\partial x}+i\frac{\partial}{\partial y} \right).$$}
\begin{equation}
\varrho_A(z,\overline{z})=-\frac{1}{\pi N}\lim_{\kappa\rightarrow 0}\partial_{\overline{z}}\partial_z \log \det H,
\label{rhoAdet}
\end{equation}
where we have introduced the $2N\times2N$ matrix 
\begin{equation}
H\equiv H(z,\overline{z};\kappa)=\left(\begin{array}{cc}\kappa \openone_N & i(z \openone_N-A) \\ i(z \openone_N-A)^\dagger & \kappa \openone_N\end{array}\right).
\end{equation}
We use $\overline{(\cdots)}$ for the complex-conjugate and $(\cdots)^\dagger$ for the conjugate-transpose. The next step is to write the determinant of $H$ in terms of a Gaussian integral.  In \cite{sommers-1988}, the replica method was applied to the case of fully connected non-Hermitian Gaussian ensembles, deriving the generalised Girko's law in the limit $N\rightarrow\infty$. However, little progress has been made in the study of sparse non-Hermitian matrices. To push forward, we tackle the problem by focusing on the behaviour of a large single instance. With a modest amount of foresight, we define $N$ pairs of complex variables,  
\begin{equation}
\psi_i=\left(\begin{array}{c}u_i\\v_i\end{array}\right) \hspace{5mm} i=1,...,N
\end{equation}
and introduce the `Hamiltonian'
\begin{equation}
\begin{split}
\sH(\bm{\psi},z,\overline{z};\kappa)=&\sum_{i=1}^N\psi_i^\dag\left[\kappa\openone_2+i(x\sigma_x-y\sigma_y)\right]\psi_i\\
&-i\sum_{i,j=1}^N\psi_i^\dag\left(A^h_{ij}\sigma_x-A^s_{ij}\sigma_y\right)\psi_j
\end{split}
\label{ham}
\end{equation}
where $\sigma_x$ and $\sigma_y$ are the usual Pauli matrices, and we have written $A=A^h+iA^s$, with $A^h$ and $ A^s$ Hermitian matrices. Continuing with the statistical mechanics metaphor, we also introduce a `distribution' $P$ and an `average' $\langle\cdots\rangle$:
\begin{equation}
\begin{split}
&P\left(\bm{\psi}\right)=\frac{1}{\sZ}e^{-\sH(\bm{\psi},z,\overline{z};\kappa)}\\
&\langle\cdots\rangle = \int \mathcal{D}\bm{\psi}\, P\left(\bm{\psi}\right) (\cdots)\,,
\label{dist}
\end{split}
\end{equation}
where $\sZ=\langle1\rangle$. Of course, the measure we define here is complex and is manifestly \emph{not} a real stochastic measure. However, much of the mathematics is unchanged, and it is relatively safe to use this probabilistic analogy. With this groundwork in place, equation (\ref{rhoAdet}) may finally be written as follows: 
\begin{equation}
\varrho_A(z,\overline{z})=\lim_{\kappa\rightarrow 0}\frac{1}{\pi N}\sum_{\ell=1}^N  i\partial_{\overline{z}}\langle \psi_\ell^\dag \sigma^+\psi_\ell \rangle,
\label{varrhoA}
\end{equation}
where $\sigma^+=\sigma_x+i\sigma_y$. In this formulation, it becomes clear that the local marginals $P_i(\psi_i)$ will suffice to evaluate $\varrho_A$. We calculate these marginals using the cavity method.\\
\indent\textit{Treelike Matrices.--} We consider treelike sparse matrices. Associated with the matrix $A$ there is a weighted, directed graph $\sG_A$ on $N$ vertices: a directed edge of weight $A_{ij}$ is drawn from vertex $i$ to vertex $j$ whenever $A_{ij}\neq0$. We say the graph $\sG_A$ (and consequently the matrix $A$) is \textit{treelike} if short loops are rare. There is a simple link between this graph and the Hamiltonian (\ref{ham}): the interaction of variables $\psi_i$ and $\psi_j$ is encoded in the edges between $i$ and $j$. We make the following standard definitions: a pair of vertices $i,j$ are neighbours if either $A_{ij}$ or $A_{ji}$ is non-zero; the set of all neighbours of $i$ is denoted $\sN i$; $k_i$ denotes the number of neighbours of $i$ (the degree of $i$); the average degree is given by $c=N^{-1}\sum_ik_i$.\\
\indent Notice that, if $A$ is treelike, the variables neighboring $\psi_i$ are correlated mainly through $\psi_i$. Consider a fictitious situation in which we have removed the variable $\psi_i$. We are interested in the change to the marginal distributions of the neighboring variables $\psi_\ell$ with $\ell \in\sN i$, which we denote by $P_\ell^{(i)}(\psi_\ell)$; with their common neighbour now absent, the joint distribution factorises:
\begin{equation}
P^{(i)}(\lbrace\psi_l\rbrace_{l\in\sN i})=\prod_{l\in\sN i} P_l^{(i)}(\psi_l).
\label{bethe}
\end{equation}
This is known as the Bethe approximation. It is exact on trees and graphs which remain treelike in the limit $N\rightarrow\infty$ \footnote{In fact, this factorisation also applies in the fully connected case, though the cause is statistical rather than topological.}. The cavity marginals $\lbrace P_i^{(j)}\rbrace$ obey simple recursive relations,
\begin{equation}
P_i^{(j)}(\psi_i)=\frac{e^{-\sH_i}}{Z_i^{(j)}}\int\mathcal{D}(\bm{\psi}_{\sN i\setminus j})\, e^{-\sum_{\ell\in\sN i\setminus j}\sH_{i\ell}}\prod_{\ell\in\sN i\setminus j} P_\ell^{(i)}(\psi_\ell),
\label{caveq}
\end{equation}
where $Z_i^{(j)}$ is a normalising constant, and we have split the Hamiltonian (\ref{ham}) into the contributions from single variables, $\sH_i$, and from pairs of variables, $\sH_{ij}$, viz.
\begin{equation}
\begin{split}
&\sH_i=\psi_i^\dagger\left[\kappa\openone_2+i(x\sigma_x-y\sigma_y)\right]\psi_i\\
&\sH_{ij}=-i\psi_i^\dagger\left(A^h_{ij}\sigma_x-A^s_{ij}\sigma_y\right)\psi_j-i\psi_j^\dagger\left(A^h_{ji}\sigma_x-A^s_{ji}\sigma_y\right)\psi_i.
\end{split}
\end{equation}
If the cavity distributions are known, the real marginal distribution at the vertex $i$ can be recovered by merging the contributions of the neighbours,
\begin{equation}
P_i(\psi_i)=\frac{e^{-\sH_i}}{Z_i}\int\mathcal{D}(\bm{\psi}_{\sN i})\,e^{-\sum_{\ell\in\sN i}\sH_{i\ell}}\prod_{\ell\in\sN i} P_\ell^{(i)}(\psi_\ell).
\label{marg}
\end{equation}
We see that the set of equations (\ref{caveq}) is self-consistently solved by distributions of a bivariate Gaussian type. Specifically, for all $i=1,...,N$ and all $j\in\sN i$, the distribution $P_i^{(j)}$ has the form:   
\begin{equation}
P_i^{(j)}(\psi_i)=\frac{1}{Z_i^{(j)}}\exp\left( -\psi_i^\dag \left[\sC_i^{(j)}\right]^{-1}\psi_i \right),
\label{gansatz}
\end{equation}
where $\sC_i^{(j)}$ is a $2\times2$ matrix. Insertion into equation (\ref{caveq}) yields a set of consistency equations for the matrices $\lbrace \sC_i^{(j)} \rbrace$, viz.  
\begin{equation}
\sC_i^{(j)}=\left[F(\sC_{\partial i \setminus j}^{(i)})+\kappa\openone_2+i(x\sigma_x-y\sigma_y)\right]^{-1},
\label{consiseq}
\end{equation}
for all $i=1,...,N$ and all $j\in\sN i$ and where $F$ is the matrix field
\begin{equation}
F(\sC_{\partial i \setminus j}^{(i)})=\sum_{\ell \in\partial i \setminus j}(A_{i\ell}^h\sigma_x-A_{i\ell}^s\sigma_y)\sC_\ell^{(i)}(A_{\ell i}^h\sigma_x-A_{\ell i}^s\sigma_y).
\end{equation}
Similarly, equation (\ref{marg}) gives the `true' covariance matrices
\begin{equation}
\sC_i=\left[F(\sC_{\partial i}^{(i)})+\kappa\openone_2+i(x\sigma_x-y\sigma_y)\right]^{-1}.
\label{true}
\end{equation}
for all $i=1,\ldots,N$. We pause for a moment now to determine the structure of the matrices $\lbrace\sC_i^{(j)}\rbrace$. Performing the inverse of $H$ in block form reveals enough information to allow us to write generally 
\begin{equation}
\sC_i^{(j)}\equiv\left(\begin{array}{cc} a_i^{(j)} & i \overline{b}_i^{(j)} \\ i b_i^{(j)} & d_i^{(j)} \end{array}\right)\hspace{7mm} \begin{array}{l}a_i^{(j)},d_i^{(j)}\in\mathbb{R}^{+} \\ b_i^{(j)}\in\mathbb{C}\end{array}.
\label{Cform}
\end{equation}
If one has a solution set to equations (\ref{consiseq}), the `true' local marginals are recovered from equation (\ref{true}). Recalling that the matrices $\lbrace \sC_i^{(j)} \rbrace$ and $\lbrace \sC_i \rbrace$, are dependent upon $z$ and $\kappa$, one may employ equation (\ref{varrhoA}) to determine the spectral density in terms of the function $b_i\equiv b_i(z,\overline{z},\kappa)$,
\begin{equation}
\varrho_A(z,\overline{z})=-\frac{1}{\pi N}\lim_{\kappa\rightarrow 0}\sum_{i=1}^N\partial_{\overline{z}}\,b_i(z,\overline{z},\kappa).
\label{spec}
\end{equation}
To deal with the partial derivative  appearing in eq. \eqref{spec}, we use eq. (\ref{consiseq}) to formulate a similar set of consistency relations for the partial derivatives of the covariance matrices, $\lbrace\partial_{\overline{z}}\sC_i^{(j)}\rbrace$:
\begin{equation}
\partial_{\overline{z}}\sC_i^{(j)}=-\sC_i^{(j)}\left[\left(\begin{array}{cc} 0 & 0 \\ i & 0 \end{array}\right)-F(\partial_{\overline{z}}\sC_{\partial i \setminus j}^{(i)}) \right]\sC_i^{(j)}.
\label{partconsis}
\end{equation}
Similarly, the derivative of the `true' covariance matrix at $i$ is given according to (\ref{true}) by
\begin{equation}
\partial_{\overline{z}}\sC_i=-\sC_i\left[\left(\begin{array}{cc} 0 & 0 \\ i & 0 \end{array}\right)-F(\partial_{\overline{z}}\sC_{\partial i}^{(i)})\right]\sC_i.
\label{parttrue}
\end{equation}
Equations (\ref{consiseq}, \ref{true}, \ref{partconsis}, \ref{parttrue}) comprise our main result. For a given treelike matrix $A$, one iterates (\ref{consiseq}) and (\ref{partconsis}) together until convergence. The `true' marginals are then recovered via (\ref{true}) and (\ref{parttrue}), and finally the spectral density is given by (\ref{spec}).\\
\indent\textit{The fully connected limit.--} To assess our approach from a theoretical viewpoint, we (re)derive the generalised Girko's law of \cite{sommers-1988} in the fully connected limit. To do so, we first rewrite $A^{s}\to v A^s$ with $v^2=(1-\tau)/(1+\tau)$. \\
\indent Consider statistically indepdendent matrices $A^h$ and $A^s$,  with 
\begin{equation}
\mathbb{E}(A_{ij}^{h})=0\quad\quad\mathbb{E}(|A_{ij}^{h}|^2)=(1+\tau)/2 c\,,
\end{equation}
and similarly for $A^{s}$. Here, $\tau$ is a parameter controlling the degree of Hermiticity; at $\tau=1$, $A$ is Hermitian, whereas, at $\tau=0$, $A$ is maximally non-Hermitian. We will study the limit of large $c$, by which we understand $k_i\rightarrow c$ and $c\rightarrow\infty$. First notice that equations (\ref{consiseq}) and (\ref{true}), along with the correlations given above, imply $\sC_i^{(j)}=\sC_i+\mathcal{O}(1/c)$. Moreover, upon introducing 
\begin{equation}
\left(\begin{array}{cc} a & i \overline{b} \\ i b& d \end{array}\right)=\lim_{c\to\infty}\frac{1}{c}\sum_{l\in\partial i }\sC_i\,,
\end{equation}
the spectral density takes the following form
\begin{equation}
\rho(z,\overline{z})=-\frac{1}{\pi}\partial_{\overline{z}} b(z,\overline{z})\,.
\end{equation}
Then, in the limit $c\rightarrow\infty$ equation (\ref{true}) yields
\begin{equation}
\left(\begin{array}{cc} a & i \overline{b} \\ ib & d \end{array}\right)=\frac{1}{ad+\left|\tau b+z\right|^2}\begin{pmatrix}
a &-i(\tau b+z)\\
-i(\tau \overline{b}+\overline{z})& d \end{pmatrix}
\label{fcl}
\end{equation}
where we have set $\kappa$ harmlessly to zero. This equation is easily solved  giving
\begin{equation}
\rho(z,\overline{z})=\left\lbrace\begin{array}{ll}\frac{1}{\pi(1-\tau^2)}&\textrm{  for }\left(\frac{x}{1+\tau}\right)^2+\left(\frac{y}{1-\tau}\right)^2\leq1 \\ 0 &\textrm{  else}\end{array}\right..
\end{equation}
This is the well known generalised Girko's law of \cite{sommers-1988}.\\
\indent\textit{Numerical Results.--} For ensembles of sparse random matrices, the cavity equations can be solved quickly by computer. We present the results, in comparison with direct diagonalisation, for  two cases: (i) symmetrically connected Poissonian random graphs with average connectivity $c$ and with asymmetric Gaussian edge weights  with zero mean and variance $1/c$ and (ii) asymmetrically connected Poissonian graphs with edge weights drawn uniformly from the circle of radius $1/\sqrt{c}$.\\
\indent Since numerical diagonalisation of large matrices is a computationally demanding task,  we have chosen to study relatively `small' matrices of size $N=1000$, even though the cavity equations are capable of handling matrices many orders of magnitude larger. In each case, we have diagonalised numerically  $10^5$ such matrices, having an overall of $10^8$ complex eigenvalues, so as to have  smooth 2d histrograms.\\
\indent For a given matrix, we use the cavity equations as a belief propagation algorithm: we iterate (\ref{consiseq}) and (\ref{partconsis}) together until convergence is reached and, then, compute the spectral density from (\ref{true}), (\ref{parttrue}) and (\ref{spec}). The results are  averaged over 1000 samples.\\
\begin{figure}
\includegraphics[width=240pt]{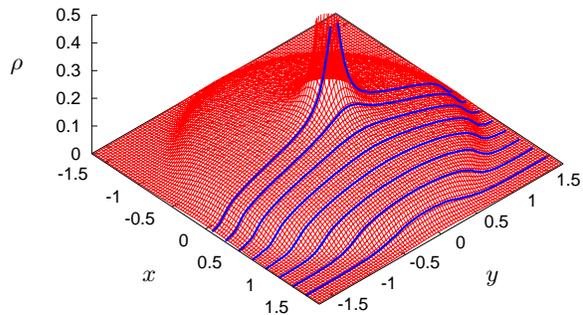}
\put(-60,30){$y$}
\put(-190,30){$x$}
\put(-240,110){$\rho$}
\caption{\label{fig:C5G} Spectral density of symmetric Poissonian graphs with asymmetric Gaussian edge weights and average connectivity $c=5$. The red grid is a histogram of the eigenvalues of $10^5$ samples, the blue lines are the result of the cavity equations, averaged over 1000 samples.}
\end{figure}
\indent The results from the cavity equations and comparison with numerical diagonalisation are presented in Figs. \ref{fig:C5G} and \ref{fig:C2U}, for cases (i) and (ii), respectively. To give a better view of the detail, Fig. \ref{fig:cuts} shows a pair of slices taken from Fig. \ref{fig:C2U}.  Notice that the ensembles in both cases satisfy the conditions for Girko's law in the limit $c\to\infty$. However, it is evident from the figures that, for finite $c$, they have spectral densities dramatically different both from each other and from the limiting case of Girko's law. Apart from small discrepances near the boundaries due to the discretization the histogram introduces, the comparison shows excellent agreement.\\
\indent\textit{Conclusions.--} In this letter we have considered the
problem of determining the mean spectral density of ensembles of
sparse non-Hermitian random matrices. Following standard steps
\cite{sommers-1988,fyodorov-1997}, the problem can be phrased in terms of a Gaussian integral,
 which we interpret in the
language of statistical mechanics of disordered systems as in \cite{edwardsjones}. Within this framework, and following the steps of \cite{rogers-2008}, we apply the
cavity method, deriving a set of equations whose solution
characterises the spectral density of a given sparse matrix.\\
\indent As we have shown, our work puts analytical results  such as the generalized
Girko's law within easy reach, when appropriate limits are taken.  Moreover, a numerical solution for
finite-size matrices can be easily obtained by belief propagation, giving results in excellent agreement with
those of direct diagonalisation.\\
\indent In the case of dense matrices (both Hermitian and
non-Hermitian), past studies using the techniques of
supersymmetry and replica analysis have found considerable success. However,
applied to sparse matrices, these approaches have not proved as
fruitful, leading to a set of saddle-point equations which have
resisted numerical solution for over 17 years.\\
\indent To make contact with these other approaches,  note that, in the ensemble average, the cavity and replica methods
are known to be equivalent. In fact, the solution we have given here
is common to all approaches and can also be derived through a careful
treatment of the aforementioned saddle-point equations
\cite{kartik}.\\
\indent For more than 50 years, rigorous analysis of the circular law has proven to be rather difficult \cite{tao2007}. Therefore, it would be a very exciting prospect to reconsider this problem by using rigorous techniques introduced in the area of spin glasses (e.g. the interpolation method). Work along these lines is under way \cite{Tim3}.

\indent \textit{Acknowledgements.--} The authors thank Reimer K\"{u}hn and Kartik Anand for discussions and Guilhem Semerjian for pointing out reference \cite{Bordenave}.
\bibliography{Asym}

\begin{thebibliography}{19}
\expandafter\ifx\csname natexlab\endcsname\relax\def\natexlab#1{#1}\fi
\expandafter\ifx\csname bibnamefont\endcsname\relax
  \def\bibnamefont#1{#1}\fi
\expandafter\ifx\csname bibfnamefont\endcsname\relax
  \def\bibfnamefont#1{#1}\fi
\expandafter\ifx\csname citenamefont\endcsname\relax
  \def\citenamefont#1{#1}\fi
\expandafter\ifx\csname url\endcsname\relax
  \def\url#1{\texttt{#1}}\fi
\expandafter\ifx\csname urlprefix\endcsname\relax\def\urlprefix{URL }\fi
\providecommand{\bibinfo}[2]{#2}
\providecommand{\eprint}[2][]{\url{#2}}

\bibitem[{\citenamefont{Guhr et~al.}(1998)\citenamefont{Guhr,
  M\"uller-Groeling, and Weidenm\"uller}}]{Guhr1998}
\bibinfo{author}{\bibfnamefont{T.}~\bibnamefont{Guhr}},
  \bibinfo{author}{\bibfnamefont{A.}~\bibnamefont{M\"uller-Groeling}},
  \bibnamefont{and} \bibinfo{author}{\bibfnamefont{H.~A.}
  \bibnamefont{Weidenm\"uller}}, \bibinfo{journal}{Phys. Rep.}
  \textbf{\bibinfo{volume}{299}}, \bibinfo{pages}{190} (\bibinfo{year}{1998}).

\bibitem[{\citenamefont{Girko}(1986)}]{Girko}
\bibinfo{author}{\bibfnamefont{V.}~\bibnamefont{Girko}},
  \bibinfo{journal}{Theor. Prob. Appl.} \textbf{\bibinfo{volume}{30}},
  \bibinfo{pages}{677} (\bibinfo{year}{1986}).

\bibitem[{\citenamefont{Bai}(1997)}]{Bai1997}
\bibinfo{author}{\bibfnamefont{Z.~D.} \bibnamefont{Bai}},
  \bibinfo{journal}{Ann. Probab.} \textbf{\bibinfo{volume}{25}},
  \bibinfo{pages}{494} (\bibinfo{year}{1997}).

\bibitem[{\citenamefont{Tao and Vu}(2007)}]{tao2007}
\bibinfo{author}{\bibfnamefont{T.}~\bibnamefont{Tao}} \bibnamefont{and}
  \bibinfo{author}{\bibfnamefont{V.}~\bibnamefont{Vu}},
  \bibinfo{journal}{e-Preprint 0708.2895}  (\bibinfo{year}{2007}).

\bibitem[{\citenamefont{Sommers et~al.}(1988)\citenamefont{Sommers, Crisanti,
  Sompolinsky, and Stein}}]{sommers-1988}
\bibinfo{author}{\bibfnamefont{H.~J.} \bibnamefont{Sommers}},
  \bibinfo{author}{\bibfnamefont{A.}~\bibnamefont{Crisanti}},
  \bibinfo{author}{\bibfnamefont{H.}~\bibnamefont{Sompolinsky}},
  \bibnamefont{and} \bibinfo{author}{\bibfnamefont{Y.}~\bibnamefont{Stein}},
  \bibinfo{journal}{Phys. Rev. Lett.} \textbf{\bibinfo{volume}{60}},
  \bibinfo{pages}{1895} (\bibinfo{year}{1988}).

\bibitem[{\citenamefont{Rodgers and Bray}(1988)}]{rodgersbray}
\bibinfo{author}{\bibfnamefont{G.~J.} \bibnamefont{Rodgers}} \bibnamefont{and}
  \bibinfo{author}{\bibfnamefont{A.~J.} \bibnamefont{Bray}},
  \bibinfo{journal}{Phys. Rev. B} \textbf{\bibinfo{volume}{37}},
  \bibinfo{pages}{3557} (\bibinfo{year}{1988}).

\bibitem[{\citenamefont{Semerjian and Cugliandolo}(2002)}]{cugliandolo}
\bibinfo{author}{\bibfnamefont{G.}~\bibnamefont{Semerjian}} \bibnamefont{and}
  \bibinfo{author}{\bibfnamefont{L.~F.} \bibnamefont{Cugliandolo}},
  \bibinfo{journal}{J. Phys. A} \textbf{\bibinfo{volume}{35}},
  \bibinfo{pages}{4837} (\bibinfo{year}{2002}).

\bibitem[{\citenamefont{Biroli and Monasson}(1999)}]{biroli}
\bibinfo{author}{\bibfnamefont{G.}~\bibnamefont{Biroli}} \bibnamefont{and}
  \bibinfo{author}{\bibfnamefont{R.}~\bibnamefont{Monasson}},
  \bibinfo{journal}{J Phys. A} \textbf{\bibinfo{volume}{32}},
  \bibinfo{pages}{L255} (\bibinfo{year}{1999}).

\bibitem[{\citenamefont{Nagao and Tanaka}(2007)}]{nagao}
\bibinfo{author}{\bibfnamefont{T.}~\bibnamefont{Nagao}} \bibnamefont{and}
  \bibinfo{author}{\bibfnamefont{T.}~\bibnamefont{Tanaka}},
  \bibinfo{journal}{J. Phys. A} \textbf{\bibinfo{volume}{40}},
  \bibinfo{pages}{4973} (\bibinfo{year}{2007}).

\bibitem[{\citenamefont{Rogers et~al.}(2008{\natexlab{a}})\citenamefont{Rogers,
  {P\'erez Castillo}, K\"uhn, and Takeda}}]{rogers-2008}
\bibinfo{author}{\bibfnamefont{T.}~\bibnamefont{Rogers}},
  \bibinfo{author}{\bibfnamefont{I.}~\bibnamefont{{P\'erez Castillo}}},
  \bibinfo{author}{\bibfnamefont{R.}~\bibnamefont{K\"uhn}}, \bibnamefont{and}
  \bibinfo{author}{\bibfnamefont{K.}~\bibnamefont{Takeda}},
  \bibinfo{journal}{Phys. Rev. E.} \textbf{\bibinfo{volume}{78}},
  \bibinfo{eid}{031116} (\bibinfo{year}{2008}{\natexlab{a}}).

\bibitem[{\citenamefont{K\"uhn}(2008)}]{Reimer}
\bibinfo{author}{\bibfnamefont{R.}~\bibnamefont{K\"uhn}}, \bibinfo{journal}{J.
  Phys. A} \textbf{\bibinfo{volume}{41}}, \bibinfo{pages}{295002}
  (\bibinfo{year}{2008}).

\bibitem[{\citenamefont{Bordenave and Lelarge}(2007)}]{Bordenave}
\bibinfo{author}{\bibfnamefont{C.}~\bibnamefont{Bordenave}} \bibnamefont{and}
  \bibinfo{author}{\bibfnamefont{M.}~\bibnamefont{Lelarge}},
  \bibinfo{journal}{e-Preprint 0801.0155}  (\bibinfo{year}{2007}).

\bibitem[{\citenamefont{Mezard et~al.}(1987)\citenamefont{Mezard, Parisi, and
  Virasoro}}]{MPV}
\bibinfo{author}{\bibfnamefont{M.}~\bibnamefont{Mezard}},
  \bibinfo{author}{\bibfnamefont{G.}~\bibnamefont{Parisi}}, \bibnamefont{and}
  \bibinfo{author}{\bibfnamefont{M.}~\bibnamefont{Virasoro}},
  \emph{\bibinfo{title}{Spin Glass Theory and Beyond (World Scientific Lecture
  Notes in Physics, Vol 9)}} (\bibinfo{publisher}{{World Scientific Publishing
  Company}}, \bibinfo{year}{1987}).

\bibitem[{\citenamefont{Mezard and Parisi}(2001)}]{mezard}
\bibinfo{author}{\bibfnamefont{M.}~\bibnamefont{Mezard}} \bibnamefont{and}
  \bibinfo{author}{\bibfnamefont{G.}~\bibnamefont{Parisi}},
  \bibinfo{journal}{Eur. Phys. J. B} \textbf{\bibinfo{volume}{20}},
  \bibinfo{pages}{217} (\bibinfo{year}{2001}).

\bibitem[{\citenamefont{{Fyodorov} et~al.}(1997)\citenamefont{{Fyodorov},
  {Khoruzhenko}, and {Sommers}}}]{fyodorov-1997}
\bibinfo{author}{\bibfnamefont{Y.~V.} \bibnamefont{{Fyodorov}}},
  \bibinfo{author}{\bibfnamefont{B.~A.} \bibnamefont{{Khoruzhenko}}},
  \bibnamefont{and} \bibinfo{author}{\bibfnamefont{H.-J.}
  \bibnamefont{{Sommers}}}, \bibinfo{journal}{Phys. Lett. A}
  \textbf{\bibinfo{volume}{226}}, \bibinfo{pages}{46} (\bibinfo{year}{1997}).

\bibitem[{\citenamefont{Feinberg and Zee}(1997)}]{feinberg-1997}
\bibinfo{author}{\bibfnamefont{J.}~\bibnamefont{Feinberg}} \bibnamefont{and}
  \bibinfo{author}{\bibfnamefont{A.}~\bibnamefont{Zee}},
  \bibinfo{journal}{Nucl. Phys. B} \textbf{\bibinfo{volume}{504}},
  \bibinfo{pages}{579} (\bibinfo{year}{1997}).

\bibitem[{\citenamefont{Edwards and Jones}(1976)}]{edwardsjones}
\bibinfo{author}{\bibfnamefont{S.~F.} \bibnamefont{Edwards}} \bibnamefont{and}
  \bibinfo{author}{\bibfnamefont{R.~C.} \bibnamefont{Jones}},
  \bibinfo{journal}{J. Phys. A} \textbf{\bibinfo{volume}{9}},
  \bibinfo{pages}{1595} (\bibinfo{year}{1976}).

\bibitem[{\citenamefont{Anand and K\"{u}hn}(2008)}]{kartik}
\bibinfo{author}{\bibfnamefont{K.}~\bibnamefont{Anand}} \bibnamefont{and}
  \bibinfo{author}{\bibfnamefont{R.}~\bibnamefont{K\"{u}hn}},
  \bibinfo{journal}{\textit{In preparation}}  (\bibinfo{year}{2008}).

\bibitem[{\citenamefont{Rogers et~al.}(2008{\natexlab{b}})\citenamefont{Rogers,
  Barra, and {P\'erez Castillo}}}]{Tim3}
\bibinfo{author}{\bibfnamefont{T.}~\bibnamefont{Rogers}},
  \bibinfo{author}{\bibfnamefont{A.}~\bibnamefont{Barra}}, \bibnamefont{and}
  \bibinfo{author}{\bibfnamefont{I.}~\bibnamefont{{P\'erez Castillo}}},
  \emph{\bibinfo{title}{In preparation}} (\bibinfo{year}{2008}{\natexlab{b}}).

\end{thebibliography}
\end{document}